\documentclass{PoS}

\title{The E\"{o}tv\"{o}s Paradox: The Enduring Significance of  E\"{o}tv\"{o}s' Most Famous Paper}
\ShortTitle{The E\"{o}tv\"{o}s Paradox}

\author{\speaker{Ephraim Fischbach}\\
Department of Physics and Astronomy, Purdue University, West Lafayette, IN 47907 USA\\
E-mail: \email{ephraim@purdue.edu}}

\author{{Dennis E. Krause}\\
Department of Physics, Wabash College, Crawfordsville, IN 47933 USA \\
Department of Physics and Astronomy, Purdue University, West Lafayette, IN 47907 USA\\
E-mail: \email{kraused@wabash.edu}}

\abstract{
Following the death of Baron Lor\'{a}nd von E\"{o}tv\"{o}s in 1919, his collaborators Desiderius Pek\'{a}r and Eugen Fekete co-authored a paper in 1922 containing the results of a series of earlier experiments testing the identity of inertial and gravitational mass, the Weak Equivalence Principle (WEP).  Although the so-called ``EPF'' paper made no claim for any WEP violations, a subsequent 1986 reanalysis of the EPF paper revealed a pattern in their data suggesting the presence of a new (``fifth'') force in nature.  Although the EPF data, and the 1986 reanalysis of these data, present fairly compelling evidence for such a fifth force, many contemporary experiments have failed to detect its presence. Here we summarize the key elements of this ``E\"{o}tv\"{o}s paradox,'' and suggest some possible paths to a resolution.  Along the way we also discuss the close relationship between E\"{o}tv\"{o}s  and Einstein, and consider how their respective contributions may have been influenced by the other's.
}

\FullConference{International Conference on Precision Physics and Fundamental Physical Constants - FFK2019\\
		9-14 June, 2019\\
		Tihany, Hungary}

\begin{document}

%===============================================================================

\section{Introduction}

	As we celebrate this year the centenary of the passing of Baron Lor\'{a}nd von E\"{o}tv\"{o}s on 8 April 1919, it is appropriate  to reflect on the experiment most closely identified with him.\footnote{A shorter version of this paper authored by Fischbach (arXiv:1901.11163) will appear as a chapter in the Roland E\"{o}tv\"{o}s Centenary Memorial Album.}  This is, of course, the work described in his last published paper: ``Contributions to the Law of Proportionality of Inertia and Gravity,'' co-authored with his collaborators Desiderius Pek\'{a}r and Eugen Fekete and published posthumously in 1922 \cite{EPF,Szabo}. As we will 
describe in detail below, an analysis of the data obtained by  E\"{o}tv\"{o}s,  Pek\'{a}r, and Fekete (EPF) published in 1986  by Fischbach, \emph{et al.} (FSSTA) \cite{FSSTA} revealed compelling evidence for the presence of a new fundamental interaction in nature, which came to be known as the  ``fifth force,'' due to a front page report in the {\em New York Times} \cite{NYT}.   However, no credible evidence for the existence of such a force has been produced to date, notwithstanding many attempts by a large number of experimental groups \cite{Adelberger,Wagner,Will}.
	
	Now that more than 30 years have passed since the publication of FSSTA, it is safe to summarize our understanding of the content of the paper by E\"otv\"os and collaborators in the following three statements: 
\begin{enumerate}
\item[\#1.]  There is consensus in the community, that there were no obvious flaws in the EPF experiment, or in their published paper.

\item[\#2.]  There is additional consensus that the analysis of the EPF data reported by FSSTA is also correct, along with its suggestion of a new ``fifth force'' in nature. 

\item[\#3.]  There is no credible experimental evidence for a new force, with the characteristics presented by FSSTA.  
 \end{enumerate}
Since the preceding three statements appear to be mutually contradictory, it is certain that interest in the EPF experiment will endure until such time that we resolve this paradox (the ``E\"{o}tv\"{o}s paradox'') with some combination of new theory and additional experiments.  In what follows we elaborate on the above observations in the hope of pointing to some possible resolutions of the E\"{o}tv\"{o}s paradox arising from the incompatibilities among observations \#1, \#2, and \#3.

\section{Correctness of the EPF Experiment}

\begin{figure}[t]
    \centering \includegraphics[width=.8\textwidth]{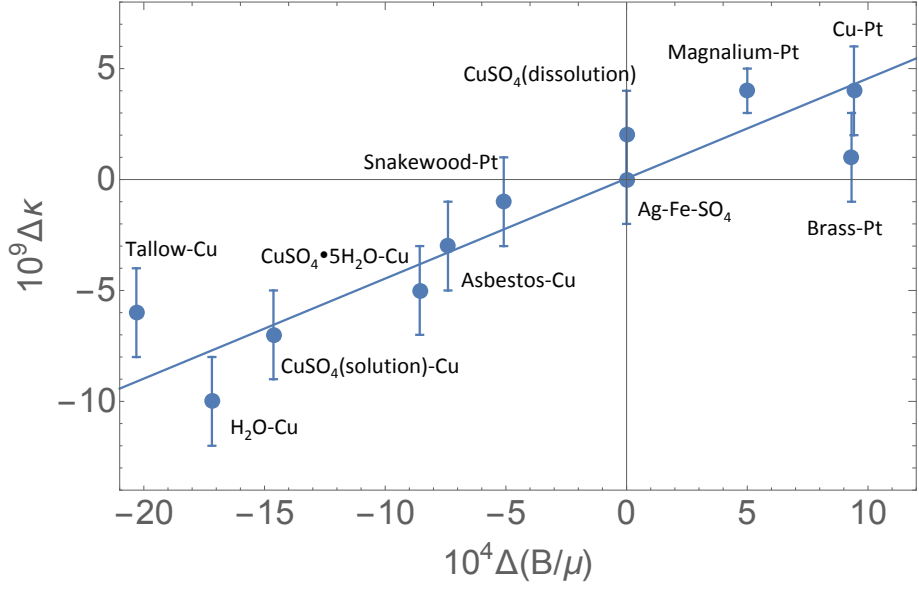}
     \caption{Plot of the data from Table~\ref{Data Table} with the best fit line from Ref.~\cite{FT-book}, revealing a dependence of $\kappa$ on the baryon-to-mass ratio.}
     \label{Data graph}
     \end{figure}

	We begin by discussing point \#1, the correctness of the EPF experiment, by briefly summarizing the results.  The goal of the experiment was to search for differences in the inertial and gravitational mass ($m_{I}$ and $m_{G}$, respectively) by examining the net torque on the arms of a torsion balance from which two different samples were suspended.  If we characterize the violation of the Weak Equivalence Principle (WEP) of a substance with the parameter $\kappa$ defined by
\begin{equation}
\kappa \equiv \frac{m_{G}}{m_{I}} - 1,
\end{equation}
then EPF measured differences in $\kappa$ for various combinations of  substances, obtaining the results given in the second column of Table~\ref{Data Table}.
\begin{table}
\begin{center}
     \begin{tabular}{|l|r|r|}  \hline
    Samples Compared & $10^{9}\Delta\kappa$ & $10^{4}\Delta(B/\mu)$ \\ \hline
    Magnalium-Pt  & $4\pm 1$  & $+5.00$ \\
    Brass-Pt & $1\pm 2$ & $+9.32$ \\
    Cu-Pt & $4 \pm 2$ & $+9.42$ \\
    Ag$\cdot$Fe$\cdot$O$_{4}$-Pt & $0 \pm 2$ & $0.0$ \\
    CuSO$_{4}$ (dissolution)-Pt & $2 \pm 2$ & 0.0 \\
    Snakewood-Pt & $-1 \pm 2$ & $-5.09$ \\
    Asbestos-Cu & $-3 \pm 2$ & $-7.40$ \\
    CuSO$_{4}\cdot$5H$_{2}$O-Cu & $-5 \pm 2$ & $-8.57$ \\
    CuSO$_{4}$ (solution)-Cu & $-7 \pm 2$ &$-14.63$ \\
    H$_{2}$O-Cu & $-10 \pm 2$ & $-17.18$ \\
    Tallow-Cu & $-6 \pm 2$ & $-20.31$ \\ \hline
       \end{tabular}
 \end{center}
     \caption{Values of $\Delta\kappa$ and $\Delta(B/\mu)$ for the various combinations of samples used in the E\"{o}tv\"{o}s experiment~\cite{FT-book} .}
     \label{Data Table}
     \end{table}
  If we simply average the values of $\Delta\kappa$ and combine the statistical uncertainty ($1.4 \times 10^{-9}$) in quadrature with the experimental uncertainties for each measurement  ($2 \times 10^{-9}$), we find
  \begin{equation}
  \left(\Delta\kappa\right)_{\rm average} \simeq (-1.9 \pm 2.4 ) \times 10^{-9},
  \end{equation}
  which is consistent with the WEP.  However, inspection of the results reveals that only 4 of the 11 values of the individual measurements of $\Delta\kappa$ agree with the WEP, with the other measured values disagreeing with the WEP by as much as $5\sigma$.  As we discuss below, these discrepancies may have contributed to the delay in the publication of these results until after E\"{o}tv\"{o}s's death.  It was not until the publication of FSSTA, when $\Delta\kappa$ was plotted versus  the baryon-to-mass ratio differences of the various substances  (Fig.~\ref{Data graph}),  that an underlying  pattern to the measurements was revealed.
  
	As noted above, following the FSSTA publication many efforts were undertaken to attribute the EPF data to conventional systematic influences, such as temperature gradients; an extensive discussion of these hypotheses is summarized in the book by Fischbach and Talmadge (FT-book) \cite{FT-book} and in Ref.~\cite{AoP}. With one possible exception \cite{Toth}, none of these has succeeded to date:  This is not surprising because the EPF data on the acceleration differences of various pairs of samples correlate with a {\it non-classical} characteristic of the samples, namely their baryonic charge-to-mass ratios $B/\mu$, where $B$ is the total baryon number of the sample, and $\mu = m/m_{H}$ is the ratio of the sample mass  $m$ to the mass of hydrogen, $m_{H} = m({\rm _{1}H^{1}}) = 1.00782519(8)u$, and $u$ is the atomic mass unit. The concept of baryon number (the number of protons and neutrons) did not arise until after the discovery of the neutron by Chadwick in 1932, many years following the completion of the EPF experiment in 1908, and its publication in 1922. The related concept of baryonic charge, for which a conservation law has been established by  Wigner in 1949 \cite{Wigner}, is clearly a quantum property.

	A dramatic example of the non-classical nature of the influence responsible for the EPF data is the comparison of their measured values of the acceleration of platinum and copper-sulfate crystals (CuSO$_{4}\cdot$5H$_{2}$O):  These two substances differ in every known conventional physical property (density, electrical conductivity, thermal conductivity, etc.). Yet, remarkably, they have very nearly the same values of the non-classical property $B/\mu$, and from the EPF data we can infer that, in fact, they have the same accelerations to the Earth as one finds from the FT-book.  
	
	Additional support for the EPF results comes from a recently discovered hand-written draft (autograph) by E\"{o}tv\"{o}s himself of what would eventually become the published 1922 paper. (A new translation of the EPF paper incorporating the E\"{o}tv\"{o}s autograph will be published as part of the E\"{o}tv\"{o}s centenary celebration.)  One of the questions that has surrounded the EPF paper since its publication has been why their results were not published shortly after the completion of their experiment in 1908.  This question becomes even more relevant in the light of E\"{o}tv\"{o}s' observation in the autograph that the sensitivity of his experiment is more than 300 times greater than that of an earlier experiment of Bessel.   In modern times, an improvement in the determination of any quantity by a factor of 300 would surely lead to immediate publication.   A possible answer to this question may be contained in the data themselves, particularly the measured fractional acceleration difference ($\Delta\kappa$ in the EPF notation) of copper and water:
\begin{equation}
\Delta\kappa(\mbox{H$_{2}$O-Cu}) = -(10 \pm 2)\times 10^{-9}.
\end{equation}
This represents a $5\sigma$ deviation from the expected null result, and hence the probability that this difference could have arisen by mere chance is 1 in 3.5 million.  To appreciate the potential significance of this datum in the present context, we note that not only did E\"{o}tv\"{o}s and Einstein know (and respect) each other, but in January 1918 Einstein even sent a booklet \cite{Booklet}  to E\"{o}tv\"{o}s about general relativity (GR).  In the covering letter accompanying the GR booklet \cite{Einstein}, Einstein thanks E\"{o}tv\"{o}s for the support for Einstein's work (specifically the WEP) by E\"{o}tv\"{o}s' research, presumably referring here to E\"{o}tv\"{o}s' earlier experiments which Einstein and Grossmann cited in their 1913 paper \cite{Einstein Grossmann}. Since this booklet discussed the importance of the equivalence principle, which seemed at variance with the water-copper datum and six others, we presume that E\"{o}tv\"{o}s must have been motivated to carefully re-examine his data.  That he stood by his results, despite their potential implications for GR, and his personal relationship with Einstein, lends strong support to our confidence that E\"{o}tv\"{o}s felt that his experiment was done correctly.  We note in passing that even prior to the critical test of GR carried by Eddington during a solar eclipse in 1919 (after E\"{o}tv\"{o}s' death), GR had already achieved a major success, by resolving a discrepancy between observation and theory for the precession of the perihelion of Mercury.
It is  not an unreasonable stretch of the imagination to speculate that the discrepancies of his data  with the WEP caused E\"{o}tv\"{o}s  to be reluctant to publish what should have been important experimental results.
	
	Other examples of the great care that E\"{o}tv\"{o}s exercised in carrying out his famous experiment are discussed in the FSSTA paper and the FT-book. Based on these references and the preceding discussion, it is safe to assume that the E\"{o}tv\"{o}s experiment, as described in the 1922 paper and the newly discovered autograph, was in fact done correctly.  Thus a resolution of the incompatibility of observations \#1, \#2, and \#3 above must depend somehow on \#2 and \#3.

\section{Correctness of the FSSTA Analysis}

	We turn next to statement \#2 dealing with the correctness of the FSSTA analysis which led to the suggestion of a ``fifth force.'' Motivated by various hints of possible deviations from the predictions of Newtonian gravity \cite{FSSTA,FF-book}, it was suggested that there existed in nature an additional long-range interaction between any two objects $i$ and $j$ which was proportional to their respective baryon numbers $B_{i}$  and $B_{j}$.    Given that $B=N+Z$, where $N$ and $Z$ are, respectively, the numbers of neutrons and protons in each object, $B$ is approximately  proportional to the mass $M$ of any object, since the magnitude of $M$ is dominated by its number of neutrons and protons.  It follows that such an interaction would behave in some ways as an additional contribution to gravity, except for presumably small deviations which would reflect differences in the actual  compositions of the interacting samples.  To give this interaction a concrete mathematical expression it was proposed that the potential energy of interaction $[V_{5}]_{ij}$ for the new interaction between the objects $i$ and $j$ had the form of a Yukawa interaction given by
\begin{equation}
\left[V_{5}(r)\right]_{ij} =  f^{2}\frac{Y_{i}Y_{j}}{r}e^{-r/\lambda}.
\label{V5}
\end{equation}
Here $f$  is a coupling constant, defining the strength of the new interaction, and $Y=B+S$ is the hypercharge quantum number, which allows $V_{5}(r)$ to also describe possible new interactions of K-mesons for which $B=0$, but which have ``strangeness'' $S \neq 0$.  (K-mesons or kaons are elementary particles discovered in the second half of 1940's, and turned out to be among the first elementary particles possessing the quantum property of strangeness. Anomalies observed in experiments involving K-mesons partially motivated FSSTA to undertake the reanalysis of the EPF experiment \cite{FF-book}.) The characteristic length $\lambda = \hbar/m_{Y}c$ accommodates the possibility that the hypothesized interaction could have a finite range ($\lambda < \infty$) if the quantum (``hyperphoton'') mediating the hypercharge interaction (analogous to the photon) had a nonzero mass ($m_{Y} \neq 0$).

Since ordinary matter has $S =0$ and always interacts gravitationally, it is straightforward to show that the combination of the Newtonian gravitational potential $V_{N}(r) = -Gm_{i}m_{j}/r$ and the potential $V_5$ characterizing the new force leads to a total interaction  potential $V(r)$ having the form 
 \begin{equation}
 V(r) = -G_{\infty}\frac{m_{i}m_{j}}{r}\left(1 + \alpha_{ij}e^{-r/\lambda}\right),
 \end{equation}
where $\alpha_{ij}= -(B_{i}/\mu_{i})(B_{j}/\mu_{j})\xi$,
$\xi = f^{2}/(G_{\infty}m_{H}^{2}),$    	                           
and $G_{\infty}$ is the Newtonian gravitational constant as $r \rightarrow \infty$, 
 
The net acceleration of the object $j$ in the force field of the Earth (object $i$) follows from the above equations. The acceleration difference of two samples of $j$ and $j'$ in the  field of the Earth is thus proportional to 
\begin{equation}
\xi \frac{B_{Earth}}{\mu_{Earth}}\left(\frac{B_{j}}{\mu_{j}} - \frac{B_{j'}}{\mu_{j'}}\right) \equiv \xi \left(\frac{B_{Earth}}{\mu_{Earth}}\right)\Delta \left(\frac{B}{\mu}\right)_{jj'}.
\label{xi}
\end{equation}
A tabulation of $B/\mu$ values for the first 92 elements in the periodic table is presented in Table 2.1 of the FT-book.   One finds that all   $B/\mu$ values are close to 1, but differ from unity at the parts per $10^{3}$.

	Remarkably, the theory which follows from the simple equations given above correctly describes the EPF data for the acceleration differences of pairs of samples $j'-j$ measured in their experiment.  As FSSTA have shown, this theory implies that the measured fractional acceleration differences $(\Delta\kappa)_{jj'}$  should be given by
\begin{equation}
(\Delta\kappa)_{jj'} = \gamma\,\Delta \! \left(\frac{B}{\mu}\right)_{jj'},
\label{delta kappa}
\end{equation}
where $\gamma$ is a constant which is determined in part by the strength of the new interaction represented by the square of the constant $f$ through the proportionality factor $\xi$.   In a pure Newtonian world where the WEP holds, $\gamma = 0$ should hold for any pair $jj'$ of samples.    However, a linear fit to the EPF data  (Fig.~\ref{Data graph}) gives a vanishing intercept, but a non-zero slope \cite{FT-book},
\begin{equation}
\gamma = (5.65 \pm 0.71)\times 10^{-6}, 
\label{gamma}
\end{equation}
which is a surprising nonzero 8$\sigma$ effect.
	
	The computations leading to this value of $\gamma$ have been described and checked in great detail elsewhere \cite{FT-book,AoP}, and hence are very likely correct given that the compositions of the EPF samples are well known.  These include water, copper, platinum, copper sulfate crystals, a copper sulfate solution, and magnalium (a magnesium-aluminum alloy).  Although snakewood (Schlangenholz) is a relatively exotic wood, the authors of Ref.~\cite{AoP} were able to obtain a sample, which was subjected to chemical analysis, from which its $B/\mu$ value was obtained.  The composition of the remaining sample, talg (fat, suet, \ldots) is somewhat uncertain, and was estimated \cite{AoP}.  Whether or not this point is included does not significantly affect the EPF (quasi-linear) correlation between $\Delta\kappa$ and $\Delta(B/\mu$)  in Eq.~(\ref{delta kappa}).

\section{Lack of Evidence of a Fifth Force}

	Summarizing to this point, it appears very likely that the EPF experiment was done correctly (observation \#1), and that the analysis of the EPF data as presented in the first FSSTA paper is also correct (observation \#2).  This leads us next to a discussion of observation \#3, that there is no experimental support for the existence of a fifth force giving rise to the deviations from the predictions of Newtonian gravity implied by \#1 and \#2.
	
	We begin by noting that $V_{5}(r)$ suggests two broad classes of tests of  Newtonian gravity: (a) composition-dependent tests such as EPF, and (b) composition-independent tests which search for an additional $r$-dependence such as might arise from $\exp(-r/\lambda)$ or some similar factor.  Even before the publication of the FSSTA-paper, composition-independent tests (also called searches for deviations from the inverse-square-law of gravity) were carried out by a number of authors.  In the post-1986 era many more such tests were carried out over various distance scales. (A detailed compilation of both composition-independent and composition-dependent tests for new interactions and related theoretical papers through 1992 can be found in Ref.~\cite{Fischbach Metrologia}.  For more recent reviews, see Refs.~\cite{Adelberger,Wagner,Will}.)
 To date the most extensive tests for composition-dependent deviations from Newtonian gravity have been carried out by Adelberger, \emph{et al.} (the E\"{o}t-Wash Collaboration, working in Seattle at the University of Washington), whose careful experiments have set stringent limits on possible deviations from Newtonian gravity over a range of distance scales \cite{Adelberger,Wagner}.  With the exception of the ``floating ball'' experiment of  Thieberger \cite{Thieberger}, no credible experiment carried out to date  suggests a deviation from the predictions of Newtonian gravity in either composition-dependent or composition-independent experiments.
	
	Having reviewed the support for our opening observations \#1, \#2, and \#3, we find that they are in fact strongly supported by a variety of experimental and theoretical results, and hence remain mutually contradictory at present.

\section{Discussion}

	Confronted with this impasse, we are forced to rethink some of the subtle assumptions that have been made in arriving to this point.   If we accept the validity of all the experimental results,  then we must re-examine the theoretical models upon which these experiments are based.   The theoretical starting point of FSSTA was based on a 1955 paper by Lee and  Yang who first raised the question of whether conservation of baryon number would lead to the existence of a long-range field in analogy to electromagnetism \cite{Lee}.    The Yukawa form of the interaction given by Eq.~(\ref{V5}) naturally arises when two particles exchange virtually a light boson, and so forms the basis of nearly all experiments searching for new forces.  However, the effect observed in the E\"{o}tv\"{o}s data is relatively insensitive to the functional form of the force acting on the test bodies.  To explain the correlation observed in Fig.~\ref{Data graph}, the force acting on the sample must be proportional to its total baryon number $B$ and must have sufficient range to interact with whatever is producing the force.  Models based on Yukawa, or simple inverse-power-law potentials, have been excluded, so we must consider more complicated models if we wish to explain the E\"{o}tv\"{o}s results while remaining consistent with all other experiments.  Recently, Mueterthies \cite{Mike} has found that the E\"{o}tv\"{o}s experiment is more sensitive than the E\"{o}t-Wash experiments for some generic force models.  Another very different set of models arises from forces due to a dark matter wind.  If dark matter interacts with baryonic matter  via a non-gravitational interaction (e.g., a ``baryonic neutrino,'' a component of dark matter  which interacts extremely weakly with baryons), it will exert a force on a test body due to the motion of the test body through the dark matter halo (e.g., Ref.~\cite{Dvorkin}).
	
Another approach to understand the EPF data is to examine unique features of the E\"{o}tv\"{o}s experiment which would make it more sensitive than other experiments.  For example, the asymmetric design of the E\"{o}tv\"{o}s apparatus (which has not been  repeated in more modern torsion balance experiments) may be important.  The recent paper by T\'{o}th \cite{Toth} describes a simple model where the EPF data may be explained by gravity gradients since the asymmetric design of the experiment makes it sensitive to such effects.  On the other hand, the asymmetry of the apparatus may also be essential to detecting new physics effects which may be averaged out or  greatly suppressed by a symmetric apparatus.  As suggested by T\'{o}th, the  best way to investigate these possibilities  would be to repeat the EPF experiment using a similar apparatus.

A perhaps more speculative approach is to return to the Guy\`{o}t experiment described in 
FT-book (p.~126) which was the precursor to the EPF experiment.  In the Guy\`{o}t experiment a pendant suspended over a pool of mercury was used to search for a difference between $\vec{g}$ (mercury) and $\vec{g}$ (pendant), where $\vec{g}$ is the local acceleration of gravity.  As noted in the FT-book, the Guy\`{o}t experiment is a direct test of the equality of the gravitational and inertial masses of the pendant and, as such depends on the Earth's rotation.  By implication so does the EPF experiment, as is also abundantly clear from EPF.  Although the Earth's rotation is clearly an influence in modern torsion balance experiments, these experiments could still produce meaningful results if the Earth stopped rotating, whereas the Guy\`{o}t  and EPF experiments would then be meaningless.  Naturally this distinction between the Guy\`{o}t/EPF experiments and the modern torsion balance experiments raises the question of whether the striking EPF data arise in some way from a coupling to $B$ which is ``activated'' or ``catalyzed'' by the Earth's rotation.  (Here we note in passing that even a modern-day version of the Guy\`{o}t experiment would likely yield a null result.  This follows from the observation that the test masses he was using, lead and mercury, are so close to each other in the periodic table that the difference in the $B/\mu$ values would have been so small, as to lead to an undetectable signal in this experiment.)
	
	We conclude with a suggestion made previously to the effect that there may be a feature of the EPF experiment which, would explain everything, but which we are ignoring because it is ``hiding in plain sight''  (Ref.~\cite{FF-book}, p.~207).  Could it be there is some characteristic of the location of the EPF experiment that we are ignoring? At the other end of the distance scale, suppose that the source of baryon number, which the EPF experiment appears to call for, is not some local feature of the Earth but is cosmological in origin.  Perhaps if we then combined the puzzle of the EPF experiment with other puzzles arising from neutrino physics, dark matter and dark energy, etc., a grand unified scheme might emerge.  It thus appears that however the paradox raised by EPF results is eventually understood, interest in that experiment will endure into the future.
	
 \acknowledgments 

	We are deeply indebted to our colleagues Virgil Barnes, Joseph Bertaux, Gabor David,  Andrew Longman, and Michael Mueterthies for many helpful conversations relating to the EPF experiment.

\end{document}